# Improved Reinforcement Learning in Cooperative Multi-agent Environments Using Knowledge Transfer

Mahnoosh Mahdavimoghaddam[1], Amin Nikanjam[1,2], Monireh Abdoos[3]

1. Faculty of Computer Engineering, K. N. Toosi University of Technology, Tehran, Iran
2. Polytechnique Montréal, Québec, Canada
3. Faculty of Computer Science and Engineering, Shahid Beheshti University, Tehran, Iran

**Abstract**

Nowadays, cooperative multi-agent systems are used to learn how to achieve goals in large-scale dynamic environments. However, learning in these environments is challenging: from the effect of search space size on learning time to inefficient cooperation among agents. Moreover, reinforcement learning algorithms may suffer from a long time of convergence in such environments. In this paper, a communication framework is introduced. In the proposed communication framework, agents learn to cooperate effectively, and also by the introduction of a new state calculation method, the size of state space has declined considerably. Furthermore, a knowledge-transferring algorithm is presented to share the gained experiences among the different agents, and an effective knowledge-fusing mechanism is developed to fuse the agents' own experiences with the experiences received from other team members. Finally, the simulation results are provided to indicate the effectiveness of the proposed method in complex learning tasks. We have evaluated our approach on the shepherding problem and the results show that the learning process has accelerated by making use of the knowledge transferring mechanism and the size of state space has declined by generating similar states based on the state abstraction concept.

## 1. Introduction

Many problems can be solved using multi-agent systems, and cooperation among agents. Implicit parallel processing in multi-agent systems speeds up the applications. Agents can cooperate together to achieve a common goal, or compete for conflicting aims and sometimes exchange or negotiate on a topic [4-6]. Reinforcement Learning (RL) is a popular Machine Learning (ML) approach for effective learning in multi-agent systems, which is employed to tackle complex decision-making problems in uncertain environments. The RL enables an agent to progressively learn a sequence of actions to maximize the returning reward from the environment that indicates achieving the desired objectives [1]. In RL, the agent carries out the act of learning through the experiences it receives throughout a period. The learning agent is assumed to be located in an unknown environment and has no prior knowledge about how to behave. The agent interacts with the environment and is rewarded based on the result of each action. The reward causes the agent to achieve optimized control by trial and error.

The process of gaining experiences in an uncertain environment is expensive and time-consuming, and the accumulated experiences must be efficiently utilized. RL methods can be widely categorized into two groups: model-based and model-free methods. Q-learning is one of the most widely-used algorithms in RL and it does not need a model of the environment to carry out learning actions. Q-Learning follows a certain policy for moving between different states by learning an action/value function. One of the strengths of this method is the ability to learn the function without having a certain model of the environment. The agent moves from one state to the next by taking an action, and in each state receives a reward corresponding to that action. The goal of the agent is to maximize the total rewards received from the environment by taking the optimal action for each state. The core algorithm consists of a simple iteration update. Using the reward value, the previous Q value is updated according to the Eq. 1.

$$Q(s,a) \leftarrow \alpha(r + \gamma V(s'))Q(s,a) \quad (1)$$

$$V(s') \leftarrow max_a Q(s',a) \quad (2)$$

In Eq. 1, $\alpha$ is the learning rate parameter and V (s′) is given by Eq. 2. The learning rate determines how much new information is preferred over an older one. The value of zero causes the agent to learn nothing, and the value of one causes the agent to validate only the new information. The discount factor, $\gamma$, determines the importance of future rewards. The zero value causes the agent to take on the greedy approach and only considers the current rewards. While the value of one causes the agent to try a longer period of time to receive a reward. If this value is more than one, it becomes a divergent factor.

One of the problems with Q-learning is the poor performance and the long time required for the learning process. It takes a long time to guarantee high performance and convergence to the optimal policy of a RL agent under all circumstances. In addition, when the agent is in a new state, its performance would be low until the learning process is complete. This may lead to the agent making the wrong decisions, thus declining the overall performance of the network. To solve this problem, it is advised to use transfer learning together with Q-learning. Unfortunately, the convergence of algorithms employing RL may only be achieved after an extensive exploration of the state-action space, which is usually highly time-consuming. One way to speed up the convergence of RL algorithms is by making use of knowledge transfer learning [3]. The type of transferred knowledge can be identified by its specificity. Low-level knowledge, such as (s, a, r, s′) instances, an action-value function (Q), a policy (π), a full task model (model), or prior distributions, could all be transferred. Knowledge mapping could be considered an indispensable part of knowledge transfer methods. The state variables and actions need to have the same semantic meaning to make the transfer process effective [3].

In this paper, we have proposed a new communication framework in which agents learn to cooperate effectively and knowledge is transferred between team members to address long convergence time problems. In addition, a state calculation method is introduced to decrease state space size considerably. We have evaluated the proposed approach on a herding problem showing that knowledge transferring mechanism will lead to a higher success rate which is defined as a percent of cows herded to the corral. Furthermore, a new state calculation method will lead to a smaller state space size. Then our approach is compared to a recent similar approach. Results reveal that our proposed framework could achieve better results.

The contribution of this paper is summarized as:

- A new communication framework is presented to achieve effective cooperation,
- A new approach for the calculation of state is proposed to reduce the size of the state space by using the concept of state abstraction,
- An efficient knowledge-transferring mechanism is introduced to accelerate the learning process.

The rest of the paper is organized as follows: Section 2 reviews the related works. The proposed learning method is presented in Section 3. Experimental results are presented in Section 4 and Section 5 is devoted to discussion. Finally, the paper is concluded in Section 6.

## 2. Related work

There have been various studies on multi-agent reinforcement learning (MARL) as a challenging research area. The MARL is employed in various domains, including robotics, resource allocation, traffic signal control, shepherding problems, and many others [4-6]. Although RL algorithms have been highly successful in multi-agent systems, the large amount of data and lengthy exploration time required by such algorithms make them intractable for many real-world robotic areas. Numerous studies have tried to address this issue such as [7, 8]. In [7], a RL and curriculum transfer learning method are introduced to regulate multiple units in StarCraft micromanagement. The principal focus of [8] is on human guidance to reduce the number of explored states and the time of learning, under the condition that each guidance message limits the set of options for action selection to a small number of actions.

Some researchers applied a knowledge-transferring algorithm to reduce the effort required for exploration and to decrease the time of learning in a large state space environment [2]. In [2] a model-based RL scheme was presented wherein the accumulated experiences are not only used for policy learning but also used for model learning to estimate the environment.

The shepherding problem was selected here and investigated. Bayazit, Lien and Amato [9] proposed a rule-based roadmap regulation system to generate homing, exploring, and shepherding swarm behaviors. Each roadmap consisted of a series of nodes joined with edges. A simple algorithm introduced by Miki and Nakamura [10] was utilized to simulate a herding task. The authors present four rules for the behavior of the swarm: cohesion to the nearest neighbor, separation to avoid collisions with obstacles or other swarm agents, run away from the shepherd(s), and random action making random stochastic movements. Similarly, the shepherd was controlled by four regulations: guidance of the flock in an objective direction, flock creation to push scattered sheep back to the group, keeping a particular distance from the flock to avoid splitting it, and cooperation to avoid shepherds overlapping. The results illustrated the algorithm considerably repeated herding behaviors [19].

A special shepherding control system was created by Razali, Meng and Yang [11]. This control system was driven by the theory of immune networks, permitting a distributed control system, and is able to adjust actively to an environment. The memory-based immune system was carried out as an analogy for the shepherding problem, where obstacles are shown as antigens, robots as B-cells, and actions as antibodies. The results of the study, while not argued in more detail, did show that the solution was not efficacious at shepherding robotic swarm agents. A follow-up research by Razali, Meng and Yang [12], however, developed a more in-depth discussion of their method. [12] introduces a refinement of the memory-based immune network that improves a robot's action-selection procedure. Strombom et al. [13] designed a simple heuristic to imitate sheep behaviors, utilizing just a single shepherd, which closely matched that of [10], where the two-dimensional (2D) simulations consisted of the same behavioral principles. Differences consisted of swarm attraction to the center of mass (CM) of their local neighborhood (in the place of just a single neighbor), and the solo-shepherd not requiring collaboration with another shepherd, and the terms guidance and flock making replaced with driving and collecting [19].

Pierson and Schwager [14] applied the same multi-shepherd regulation strategy, where the shepherds steered a swarm agent by spreading themselves along an arc centered on the agent, using a single continuous control law, which negated the requirement for behavioral switching. Fujioka and Hayashi [15] assessed a substitute shepherding control behavior called V-formation control, in which the shepherd cycles among three positions along an arc extending out from a swarm's CM, centered behind, and rotated slightly to the left and right ($< 90°$ from the center position), making a V-shaped trajectory. Another algorithm, presented by Lee and Kim [16], applied an alternative set of swarm and shepherd behaviors to come up against a shepherding problem. If swarm agents are within a specific distance from each other, then they endeavor to minimize the distance between themselves. The agents then try to move to the center of their swarm neighborhood, while striving to match their neighbors' velocities. The swarm agents keep away from collisions with both obstacles and the shepherds. The shepherds collect scattered agents into a single swarm, and then guide them towards a goal destination. Lee and Kim, then, included patrolling (keeping the swarm at a fixed position) and covering (steering the swarm to multiple goals) shepherd behaviors into their simulations. An extra methodology and discussion were added, and their algorithm was shown to be useful at independently regulating the swarm [19]. In 2018, Hoshi et al. performed two studies [17], [18], which elaborated on the work of Strombom et al. [13]. The first in [17] tested the effects of differing the step size per time step of the shepherd and swarm agents on the shepherding task. The task was carried out better once the shepherd was able to move faster in comparison with a swarm, which moved at the same pace. In the next study [18], the shepherd and swarm agent's movement and influential force vectors were added to the third dimension [19].

Among all the research that has been done in the field of herding, a research gap in the field of reducing the size of the state space and improving the coordination among the agents is quite visible. Reducing the number of explored states and the learning time by restricting the set of choices for action selection to a small number of actions inevitably reduces the independence of the agent. Using the concepts of knowledge transfer for increasing the speed of learning and reducing the time of convergence can have a significant impact on the learning algorithms used in the mentioned research.

## 3. The proposed approach

In this section, we introduce a new communication framework for the herding problem. In the native MASSIM implementation[1], the state is calculated based on coordinates (x-y axis). If the state is calculated based on x-y axis coordinates, the size of state space would be very large as the environment is a 100x100 grid; therefore, introducing a new method for calculating the state is essential. Another point worth mentioning is that in the original MASSIM implementation, all agents must explore the environment extensively to find the best policy. However, in our proposed framework, the knowledge is transferred between agents, declining the effort required for exploration.

### 3.1. Problem

We have used the herding problem introduced in the Multi-Agent Programming Contest[2] in 2008. In this version, agents in a team can move from one cell to a neighboring one. The scope of the competition includes two teams of agents collecting cows scattered around the world and herding them into a corral. As it can be seen in Fig. 1, the green, yellow, light gray, and dark areas are tree, corral, opponent's corral, and unknown, respectively. The white color represents the cows and the two colors red and blue demonstrate the two rival teams.

---

[1] https://multiagentcontest.org/2008/scenario-r4.220.pdf
[2] https://multiagentcontest.org/2008/

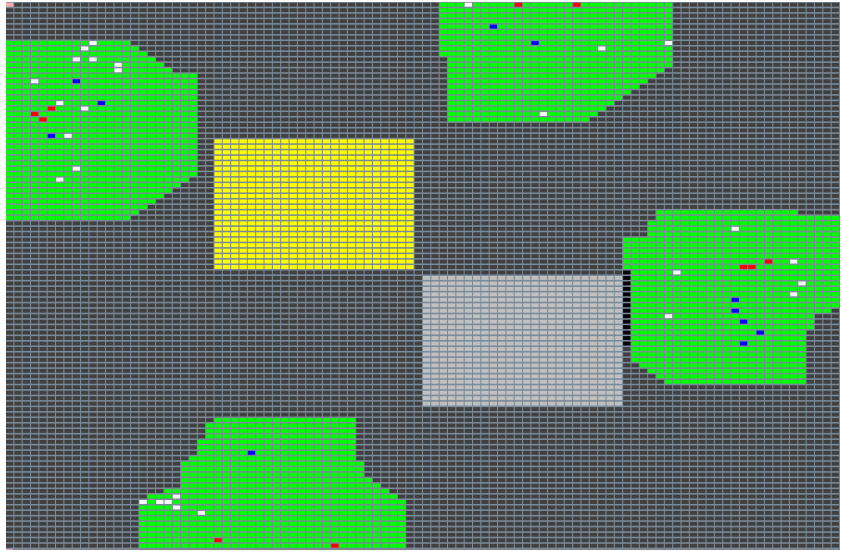

Fig. 1: Graphic space of the shepherding problem

The game server (MASSIM) interacts with player agents via TCP/IP sockets employing an XML-based messaging protocol. Within a game, the server issues messages to agents explaining their current perceptions and permitting every agent to submit an action during a certain period of time. Each agent of the team receives an XML message containing the exact position of cows, boundaries, other agents, and corral at each time step. Each agent or group of agents tries to steer the surrounding cows or even steal them from the opponent team to make the flock size bigger and bigger. Then each agent or group of agents collect cows and push them into the corral. The following scenarios are repeated for all agents in each timestamp:

1. Understanding the grid (finding cows, boundaries, other agents, and corrals),

2. Engaging with the opponent team (Stealing the opponent's collected cluster in the middle of their path),

3. Collecting cows and pushing them into our corral.

The original implementation of the system is done in the JIAC[3] programming language, which is a Java extension. JIAC provides a Distributed Communications Infrastructure (DCI), where each agent is responsible for its processing and functionality. Therefore, decentralization makes robotic systems easier to build and program, as well as introduces robustness, flexibility, and scalability into the multi-robot system. All agents receive XML messages from the game server corresponding to the various game phases: the beginning of a game (sim-start); the end of the game (sim-end); the end of a simulation (bye); and within a game, where the game server constantly requests a next move (request-action). The system sends action demands to the game server; either to move a player to an adjacent cell, or a "skip" action, i.e., no movement (the game server additionally implements a response timeout, defaulting the action to be a skip action if the system fails to issue its action choice during the time limit).

---

[3] https://www.aot.tu-berlin.de/fileadmin/files/lehre/WiSe_06/IV_AOT/Uebung/Dokumente/JIAC-ProgrammersGuide.pdf

We have chosen the herding problem in this paper due to the following items:

1. Herding agents want to cooperate to herd the cows to the corral since effective cooperation maximize the team reward and achieve a high percentage of success (a large number of cows in the corral),

2. If the state is calculated based on x-y axis coordinates, the size of state space would be very large as the environment is a 100x100 grid, therefore, introducing a new method for calculating the state is essential,

3. Exploring the whole environment would be highly time-consuming; however, knowledge transfer will address this issue. It is worth mentioning that the new representation of state paves the way for knowledge transfer as there are many similar states.

### 3.2. System roles

The system consisted of a single coordinator agent and several player agents. While the former is responsible for team-level support functionality, the latter are those playing the actual game on the game server. In the communication framework, the coordinator agent was introduced as an internal agent to the system, as it does not interact with other agents in the environment.

The MASSIM server interaction is described as a perception + act cycle, where the agent in every time step receives a percept and submits an action. The player agent type will represent agents registered in the game server and play the actual game. These are the agents that can perceive and act in the game server. A coordinator agent can fulfill the coordination functionality, which means facilitating team behavior within a group of player agents.

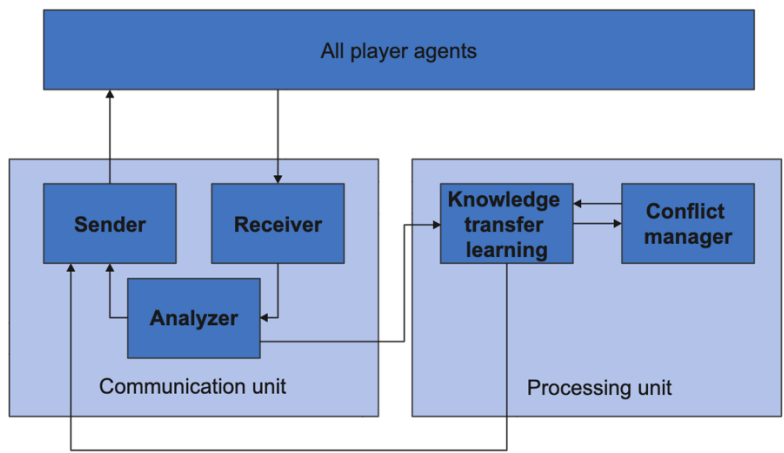

Fig. 2: Structure of the coordinator agent

At the beginning of the simulation, the server picks two teams of agents to participate in the simulation. The simulation starts by notifying the corresponding agents about the details of the starting simulation. This notification is done by sending the SIM-START message. This message contains the upper, lower, left, and right border of the corral (numeric)[4].

According to Fig. 2, in the first step of the game, all player agents send their coordinates to the coordinator agent. As this is not an obstacle-free environment, the coordinator agent is responsible for finding the player agent that is closer to the middle of the corral. Then, a message will be sent to the closer agent to make it responsible for identifying valid corral entrances. After all the entrances have been identified, this agent sends the array of valid corral entrances coordinates to the coordinator agent, and the coordinator agent sends this array to all player agents.

The second function of this agent is related to the concept of knowledge transfer. At the end of each time step, all player agents send their Q-tables to the coordinator agent. The receiving unit, after receiving the Q-tables, sends them to the analyzer unit. If the received message was a table type, the analyzer unit will send that table to the knowledge transfer unit. In the conflict manager unit, the knowledge of all similar Q-tables (each behavior has a separate Q-table) is aggregated into a table and this table is returned to the knowledge transfer unit to be sent to all agents. The functionality of the Analyzer unit is shown in Algorithm 1.

**ALGORITHM 1: ANALYZER UNIT**

| | |
|---|---|
| 1 | *Initialization of variables: Counter1←0, Counter2←0, Closer←agent[0].id, **Q-tables** ← Array()* |
| 2 | *GetMessage(**Message** message)* |
| 3 |     **if** *message.type = 'coordinate'* **then** |
| 4 |         *Counter1++* |
| 5 |         **if** *is_Closer(message.coordinate, agent[closer].coordinate)* **then** |
| 6 |             *Closer ← message.id* |
| 7 |         *end* |
| 8 |         **if** *Counter1 = Number_of_agents* **then** |
| 9 |             *Sending a message to the agent closer to the middle of the corral to identify corral entrances* |
| 10 |         *end* |
| 11 |     *end* |
| 12 |     **else if** *message.type = 'entrances'* **then** |
| 13 |         *Sending a message containing all entrances of the corral to all agents* |
| 14 |     *end* |
| 15 |     **else if** *message.type = 'Q-table'* **then** |
| 16 |         *Counter2++* |
| 17 |         *Aggregate all the same Q-tables into a single Q-table based on equation(5)* |
| 18 |         **if** *Counter2 = Number_of_agents* **then** |
| 19 |             *Sending all the same single Q-tables to all agents* |
| 20 |             *Counter2←0* |

---

[4] https://multiagentcontest.org/2008/protocol.pdf

| | | |
|---|---|---|
| 21 | | end |
| 22 | | end |
| 23 | end function | |
| 24 | is_Closer (point1,point2) | |
| 25 | | *Distance1*←*Euclidean distance between **point1** and middle of the corral* |
| 26 | | *Distance2*←*Euclidean distance between **point2** and middle of the corral* |
| 27 | | *if Distance1 < Distance2 then* |
| 28 | | *Return true* |
| 29 | | *end* |
| 30 | | *else Return false* |
| 31 | end function | |

In the beginning, all player agents are waiting to receive a message from the coordinator agent to start searching around the corral. Random action is selected before receiving that message. After the closer agent to the middle of the corral receives the message, the other player agents continue to select the random action. However, the closer agent selects actions according to the upper, lower, left, and right border of the corral (numeric) that have been received from the SIM-START message. In other words, the closer agent will be responsible for finding obstacle-free entrances to the corral. After all the valid entrances have been found, the array of valid entrances coordinates will be sent to the coordinator agent through the communication unit, and the coordinator agent sends this array to all the player agents.

After specifying all the valid entrances, the process of training the agents will start. In each time step, the nearest corral entrance to the collected cows was considered as the target. Fig. 3 shows the structure of the player agents. Observations received from the environment are delivered to the percept analysis unit. The player agent will then follow the following steps.

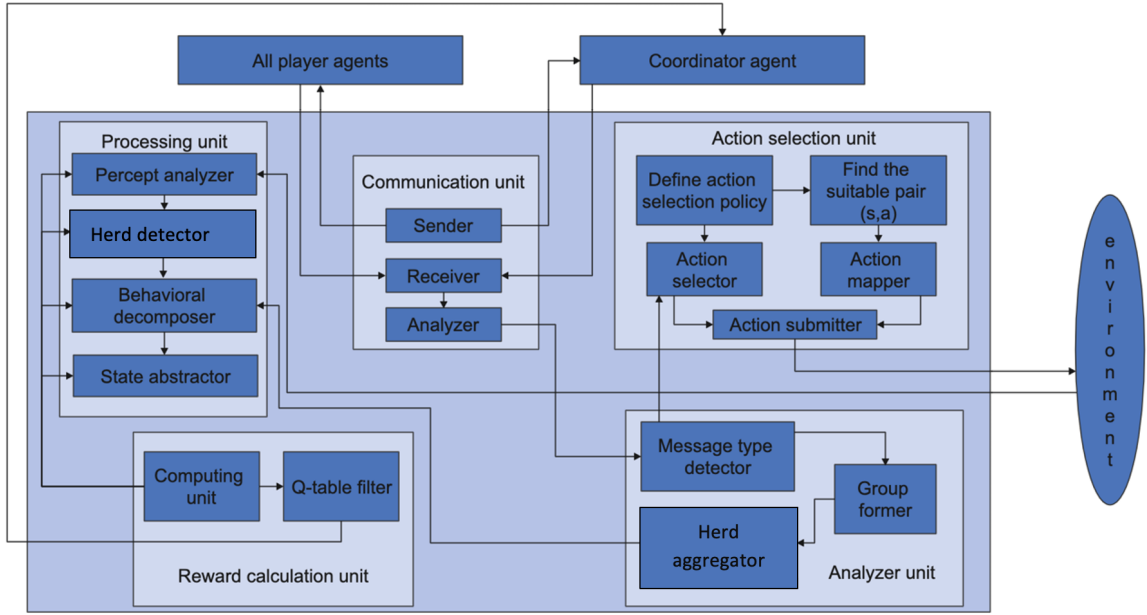

Fig. 3: The structure of the player agent

Each agent receives the information of adjacent four 8×8 squares (the line of sight is 8) and then performs the following tasks in the herd detector unit:

- Sheep clustering based on proximity threshold
- Selecting a more populous cluster to be steered towards the corral

If two or more agents want to cooperate with each other, the populous cluster calculated separately for all cooperating members will be aggregated in the herd aggregator unit.

As mentioned earlier, in each time step, the nearest valid corral entrance to the collected cluster was considered as the target. Then, the environment was divided into two parts, face the target and back with the target. According to Fig. 4, if the target point is on the right border of the corral, the environment is divided into these two parts, *A* and *B*.

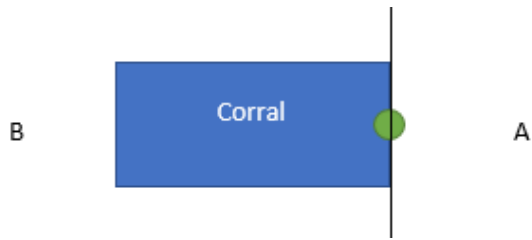

Fig. 4: Segmentation of the environment around the corral

Inspired by the model presented by Strömbom et al [13], shepherding requires two basic behavioral building blocks and the switching between them, namely, collecting and driving. Based on this definition, the following behaviors are introduced in the behavioral decomposer unit:

1) **Solo herding**

If the single agent and the collected cluster are in zone *A* and the angle between the agent and the collected cluster and the target point is greater than a threshold (greater than $120 \pm 10$).

**2) Group herding**

If the multi-agent and the collected cluster are in zone *A* and the angle between the agent and the collected cluster and the target point is greater than a threshold (greater than $120 \pm 10$).

**3) Solo following**

If the single agent and the collected cluster are in zone *A* and the angle between the agent and the collected cluster and the target point is less than a threshold (less than $120 \pm 10$).

**4) Group following**

If the multi-agent and the collected cluster are in zone *A* and the angle between the agent and the collected cluster and the target point is less than a threshold (less than $120 \pm 10$).

**5) Solo transferring**

If the single agent and the collected cluster are in zone *B*.

**6) Group Transferring**

If the multi-agent and the collected cluster are in zone *B*.

The shepherding behavior is deconstructed into switching, solo or group herding, solo or group following, or solo or group transferring. Each of these six behaviors has its own separate Q-tables.

RL depends on the state of the agent in each time step. If a state is calculated based on the x-y axis coordinates, the size of state space would be very large. By using state abstraction, irrelevant features of state space can be ignored and the size of state space can be reduced. In the State Abstractor unit, a new calculation for a state is presented. According to Fig. 5, when agent *i* is alone, the state in each time step could be described as in Eq. 3:

$$S_{a_i} \simeq \left(\left[\frac{D_{a_i t}}{d}\right], \left[\frac{D_{a_i c}}{d}\right], \left[\frac{\alpha}{a}\right]\right) \quad (3)$$

In Eq. 3, R is the size of each side of the environment as well as $1 \ll d \ll R\sqrt{2}$ and $1 \ll \alpha \ll 180$. $D_{a_i t}$ is the distance between agent ith and target. $D_{a_i c}$ is the distance between agent $i^{th}$ and GCM (global center of mass). $\alpha$ is the angle between agent $i^{th}$ and GCM and target.

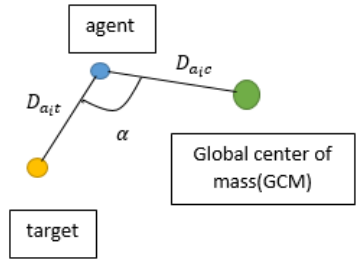

Fig. 5: One-agent state

As you can see in Fig. 6, when the number of agents pursuing the goal is more than one the state could be described as in Eq. 4:

$$S_q = \cdots = S_p \simeq \left( \left[\frac{\sum_{i=q}^{p} D_{a_i t}}{M \times d}\right], \left[\frac{\sum_{i=q}^{p} D_{a_i c}}{M \times d}\right], \left[\frac{\sum_{i=q}^{p} \alpha_i}{M \times a}\right] \right) \quad (4)$$

In Eq. 4, $R$ is the size of each side of the environment as well as $1 \ll d \ll R\sqrt{2}$ and $1 \ll \alpha \ll 180$ and $i \in \{q, \ldots, p\}$ and $M$ is the number of agents that are in the same group. By removing features unrelated to the joint action and using the concept of action abstraction, the joint action could be considered as the sum of the actions of all the agents of a group.

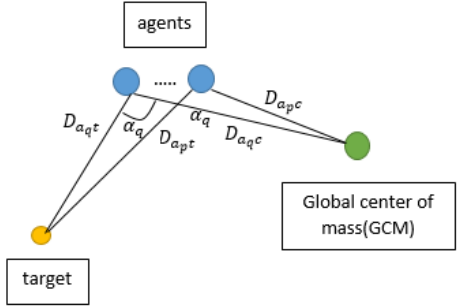

Fig. 6: Multi-agent state

Since the environment is continuous, through factors d and a, many states are similar to each other and can be stored as one state, thus significantly reducing the size of the search space.

In the action selection unit, based on the action selection policy, either action is selected randomly or is selected according to the appropriate pair(s, a). Finally, the action is submitted. At the next time step, in the reward calculation unit, the Q-table corresponding to the previous behavior of the agent will be updated according to the reward function and the changes in the perceptions. In Algorithm 2 functionality of the player agent has been demonstrated.

**ALGORITHM 2: PLAYER AGENT FUNCTIONALITY**

1 *Initialization of variables: Detector←0, Entrances←Array(), Q-table-solo-herding←Array()*

2 *Q-table-group-herding←Array(), Q-table-solo-following←Array()*

3 *Q-table-group-following←Array(), Q-table-solo-transferring←Array()*

```
4   Q-table-group-transferring←Array(), Total-iteration←50000
5   Step←ExtractFromXmlMessage(step), Action← RandomAction()
6   Herd←Array(), Behavior←'', State←''
7   Exploration← IsExploring()
8   if Step = 1
9       Send coordinates to the coordinator agent
10  end
11  else if Step<Total-iteration
12      if Detector = 1 then
13          if !FullRound() then
14              Action←take action to explore around the corral and fill array entrances
15          end
16          else
17              Detector←0, Send an array of Entrances to coordinator agent
18          end
19      end
20      else
21          herd← detecting herd, Behavior← decomposing behavior
22          State← state identification,
23          Behavior-related Q-table for the previous state is going to be filled
24          if exploration then
25              Action← RandomAction()
26          end
27          else
28              Action← MapAction(GetAction(pair(s,a)))
29          end
30          Submit Action
31      end
32  end
33  GetMessage(Message message)
34  if message.type = 'closer' then
35      Detector←1
36  end
37  else if message.type='entrances' then
38      Entrances←message.entrances
```

```
39  3  end
40  4  else if message.type='Q-table' then
41  4         Assign message.Q-table based on message.behavior to different Q-tables
42  4  end
43  4  else if message.type='cooperation' then
44  4         herd←aggregate herds of those agents wanting to cooperate with each other
45  4  end
46  4 end function
```

Function FullRound(), returns true if the agent who is responsible for detecting available entrances, completes a full lap around the corral. Function IsExploring(), establishes trade-off between exploration and exploitation.

### 3.3. Knowledge transfer

In the proposed cooperative knowledge transferring, each agents' policy/model is shared with the others, and an agent has two important tasks [2]:

- Effectively combine the newly obtained knowledge from the other agents with the knowledge gained using its own experience
- Effectively map the shared knowledge to the existing state

In this approach, each agents' Q-table is shared with the other teammates. The main task for an agent is to effectively combine the newly obtained knowledge from the other agents with the knowledge gained using its own experience. For this purpose, at the end of each time step, all agents send their Q-tables to the coordinator agent. The receiving unit, after receiving the Q-tables, sends them to the analyzer unit. If the received message was a table type, the analyzer unit will send that table to the knowledge transfer unit. In the conflict manager unit, the knowledge of all similar Q-tables (each behavior has a separate Q-table) is aggregated into a table and sent to all player agents.

In the conflict manager unit, each Q-value in the new table could be described mathematically as in Eq. 5:

$$Q_{SH}(s,a) = \frac{\sum_{i=1}^{N} Q_{SH_{a_i}}(s,a) \times M_{SH_{a_i}}(s,a)}{\sum_{i=1}^{N} M_{SH_{a_i}}(s,a)} \quad (5)$$

In Eq. 5, N is the number of agents. The above formula shows that the Q-value for the pair s and a in the solo herding table is the weighted average of the Q-value in all agents. The $M_{SH_{a_i}}(s,a)$ shows the number of times pairs s and a occur for agent i in solo herding behavior where $1 \leq i \leq N$. The way of calculation in Eq. 5 is effective as the role of each agent in determining the final Q-value is directly related to the number of times pair (s, a) occurred in an agent.

If the action is selected from Q-table, it needs to be mapped to an action which is appropriate to the current situation because this action is the result of experiences of other teammates and must be mapped to be useful in the new situation. As can be seen in Fig. 3, the action mapper unit is responsible for mapping the action.

As can be seen in Fig. 7, imagine a scenario in which agent A performs the northeast action and updates its solo following Q-table. Then, all solo following Q-tables are aggregated and a single one is sent to all agents by the coordinator agent. Assume that agent B is in the same state with agent A and wants to use agent A's experience. If it chooses the northeast action, the experience is not transferred correctly. The need for a mapping function that maps the action of a Q-table to action appropriate to the current situation is now clearly felt.

There is a model for predicting the movement of cows[5]. The action is considered as the difference between the current state and the predicted one and the action making the closest changes is selected in the action mapping unit.

**3.4. Heuristic algorithms**

When using RL approaches, the use of heuristic algorithms will increase speed and improve efficiency. We utilize rotational motion and middle action heuristic algorithms.

The rotational motion algorithm consists of agent(s) that are in the behavior of solo or group following. They choose an action that puts them in the right position behind the herd to guide the herd to its destination easily instead of acting randomly.

The middle action algorithm consists of agent(s) that are in the behavior of solo or group herding. They choose an action whose direction is closer to the direction of the line between GCM (global center of mass) and destination instead of acting randomly.

The efficiency of the agent's knowledge transfer mechanism was evaluated by comparing success rates when using heuristic algorithms and knowledge transfer.

---

[5] https://multiagentcontest.org/2008/scenario-r4.220.pdf

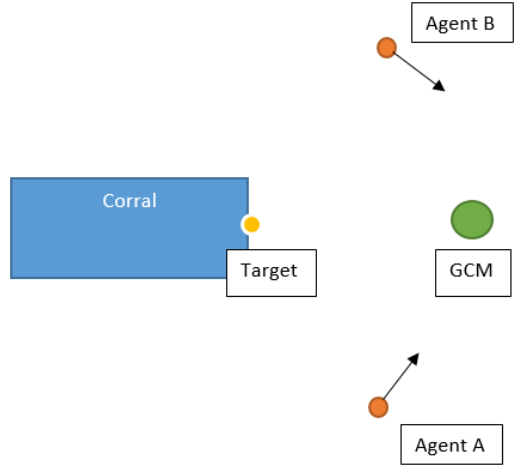

Fig. 7: Effectively map

## 4. Experimental results and analysis

In this section, we present our experimental results. Each game will run 50,000 iterations. We tried to examine the effect of knowledge transfer and the way of calculating the state by simulating the game in different situations. The environmental conditions and the size of the state space were changed according to the tables below:

Table 1: Three different environmental conditions

| Scenario 1 | 130 cows and 160 obstacles |
|---|---|
| Scenario 2 | 400 cows and 160 obstacles |
| Scenario 3 | 400 cows and 600 obstacles |

Table 2: Two different conditions for state space

| State space 1 | $d=10$ and $a=5$ |
|---|---|
| State space 2 | $d=20$ and $a=10$ |

$R$ is the size of each side of the environment as well as $1 \ll d \ll R\sqrt{2}$ and $1 \ll a \ll 180$.

Criteria for evaluating knowledge transfer are:

- Jumpstart: the rate of improvement of basic efficiency when using knowledge transfer compared to when knowledge transfer was not used. This number could be concluded based on the diagram

- Transfer rate: the rate of total rewards collected when using knowledge transfer could be described mathematically as in Eq. 6:

$$R = \frac{S_{withtranster} - S_{withouttransfer}}{S_{withouttransfer}} \quad (6)$$

In Eq. 6, $S_{withtranster}$ is the area under the curve while using the concept of knowledge transfer and $S_{withouttransfer}$ is the area under the curve while not using the concept of knowledge transfer. For understanding the Figures, the definition of success needs to be defined which is the percent of cows herded to the corral in each time step.

Our framework is based on knowledge transferring and uses a reward that is defined as a percent of cows herded to the corral. According to Fig. 8-11, the gained reward using our framework reaches its maximum value. This means that agents belonging to the same team are attempting through their interaction, to maximize reward. In our proposed framework agents belonging to the same team make use of the transferred knowledge and achieve higher rewards in comparison to when the knowledge transferring method is not used. Hence, one can conclude that agents cooperate effectively in this framework. In the following, we will discuss the results in detail.

In Figs. 8 and 9, we present the results of experiments conducted to explore the effect of d and a value on the success rate. When $d = 10$ and $a = 5$ the size of state space is equal to 7056 and when $d = 20$ and $a = 10$ the size of state space is equal to 882 and if the concept of state abstraction was not used, it will be equal to 80000. As can be seen, in both cases, there is a significant reduction in the size of the state space. When the values of $d = 20$ and $a = 10$ were used, better results were obtained because the number of similar cases increased and that enhanced the performance of the RL algorithm. The values of $d = 20$ and $a = 10$ were used to examine the effect of knowledge transfer because increasing the number of similar cases has a greater effect on knowledge transfer.

According to Fig. 9, when 130 cows are scattered in the environment and the number of obstacles is 160, the Jumpstart is about 29% when using knowledge transfer and 14% when not using knowledge transfer and the knowledge transfer rate is equal to 0.414. As can be seen in Fig. 10, when 400 cows are scattered in the environment and the number of obstacles is 160, the Jumpstart is about 66% when using knowledge transfer and 24% when not using knowledge transfer and the knowledge transfer rate is equal to 0.598. As can be seen, knowledge transfer has increased the speed of convergence as well as increased efficiency.

As can be observed in Fig. 12, the effect of knowledge transfer is more significant when the number of cows scattered in the environment is greater, and the rate of knowledge transfer is greater when the number of cows in the environment is higher. In Fig. 11, the number of obstacles was almost quadrupled. When 400 cows are scattered in the environment and 600 obstacles are in their way, the Jumpstart is about 42% when using knowledge transfer and 24% when not using knowledge transfer and the knowledge transfer rate is equal to 0.312.

As can be seen in Fig. 13, when the number of obstacles is quadrupled, good performance through knowledge transfer is achieved. The performance of the knowledge transfer mechanism could be examined in Fig. 14. While heuristic functions are used, the actions are selected correctly. By considering the proximity of the two charts, it can be concluded that the action mapping function works properly. In Table1 the summary of Fig. 8-11 could be seen.

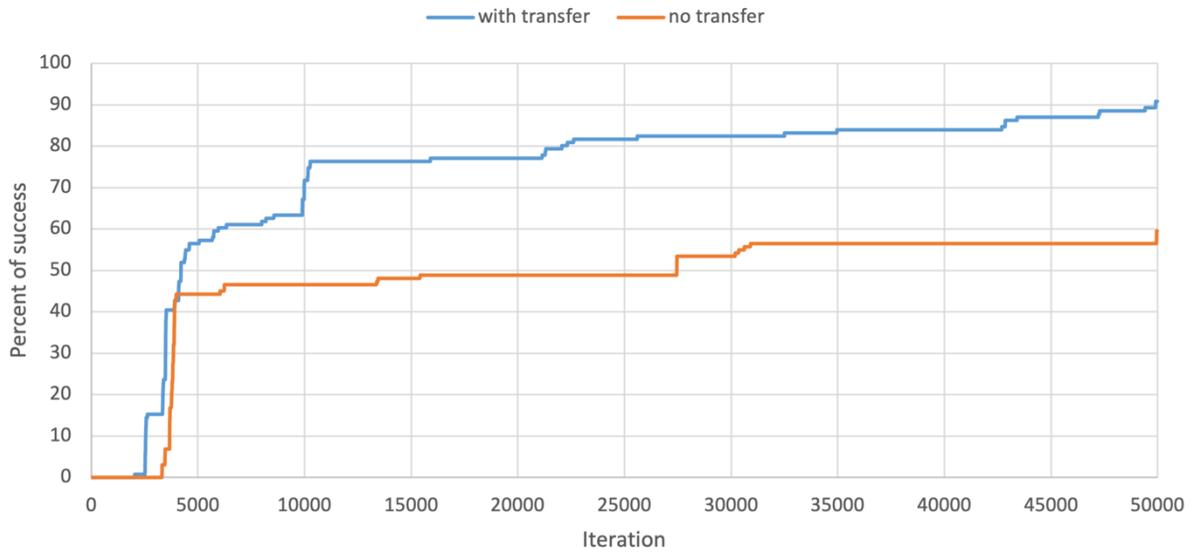

Fig. 8: Scenario1 with state space 1

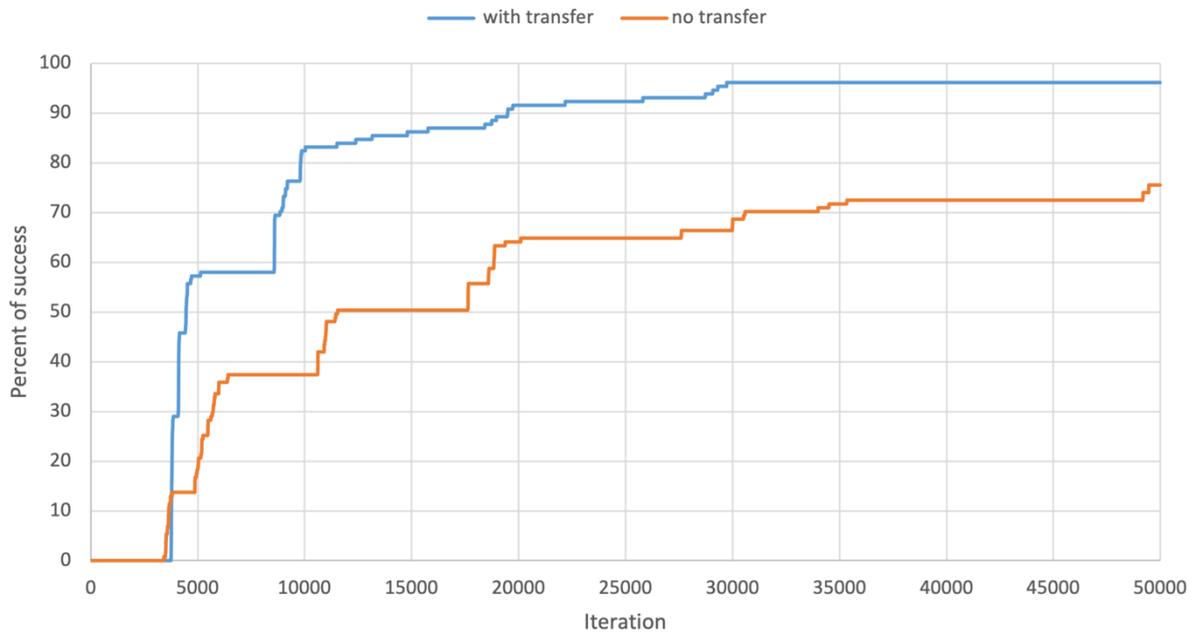

Fig. 9: Scenario1 with state space 2

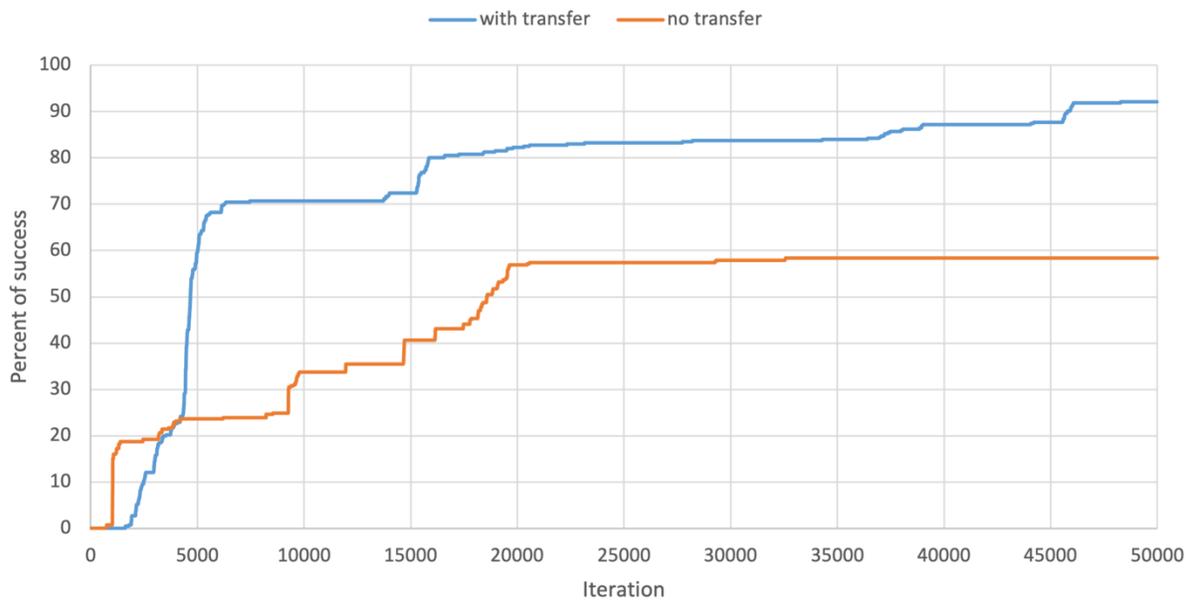

Fig. 10: Scenario 2 with state space 2

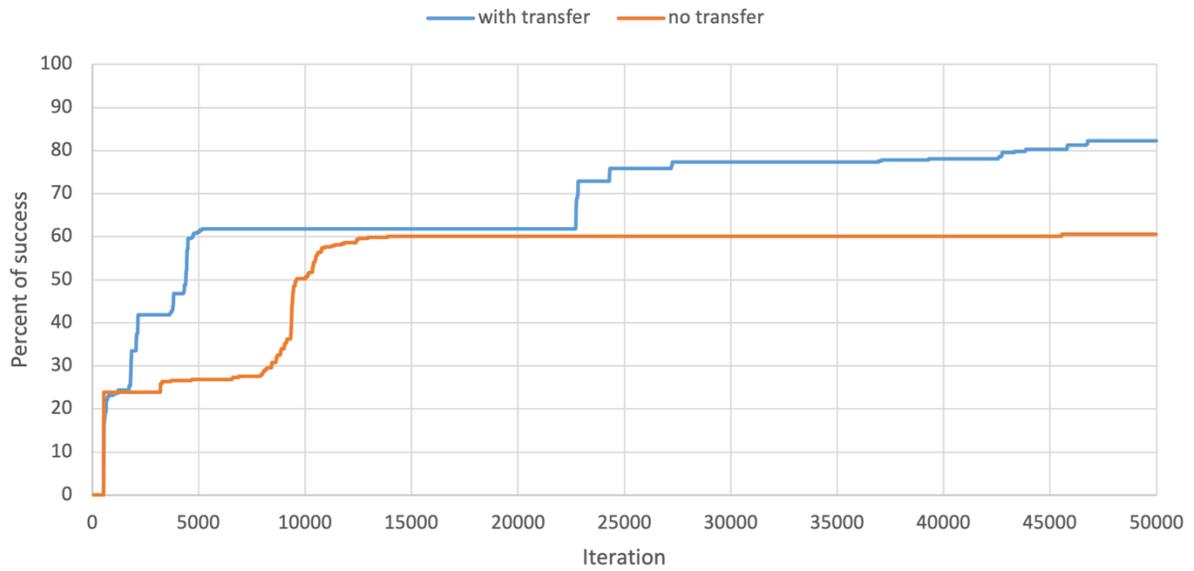

Fig. 11: Scenario 3 with state space 2

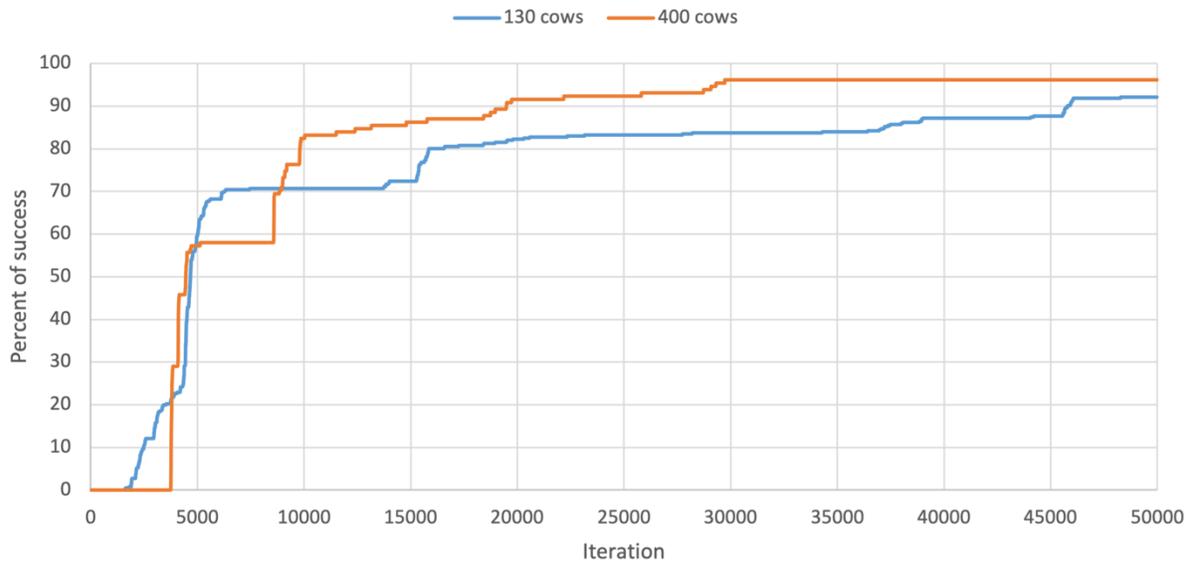

Fig. 12: Comparison between scenario 1 and scenario 2

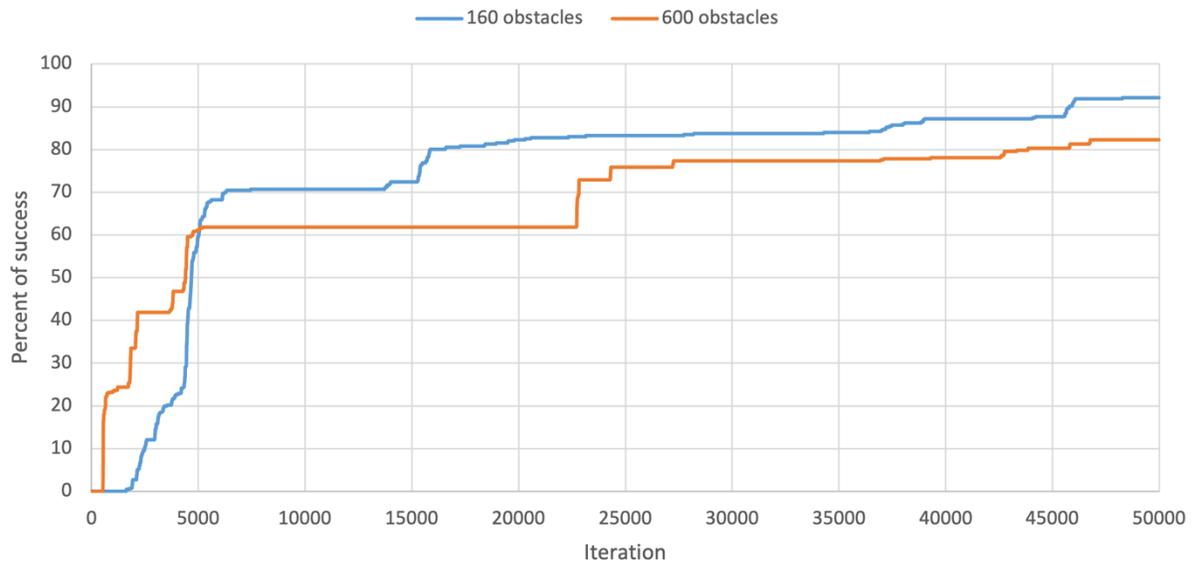

Fig. 13: Comparison between scenario 2 and scenario 3

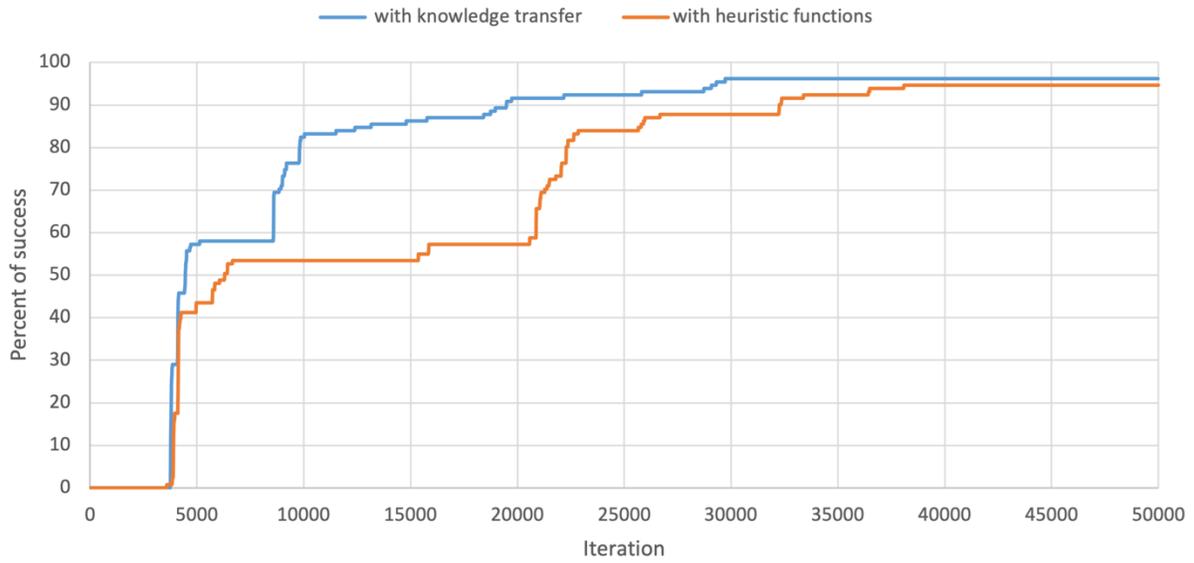

Fig. 14: Scenario1 with state space 2

Table 3: Summary of three Scenarios with two state spaces

| Condition | Scenario | State space | Success (%) | Convergence iteration |
|---|---|---|---|---|
| With transfer | One | One | 90.83 | 49913 |
| | | Two | 96.18 | 29726 |
| | Two | Two | 92.11 | 48278 |
| | Three | Two | 82.26 | 46764 |
| No transfer | One | One | 59.54 | 49958 |
| | | Two | 75.57 | 49458 |
| | Two | Two | 58.37 | 32574 |
| | Three | Two | 60.59 | 45553 |

It could be concluded that our proposed communication framework achieves a high rate of success (96.19) and this demonstrates that agents cooperate effectively. With the introduction of a new state calculation method, the size of state space declines. To be precise, the size of state space while using state space 1 and state space 2 is 7056 and 882 respectively. This reduction of size makes our Q-table smaller and has no negative effect on the rate of success (percent of cows in the corral).

We have introduced a framework in which:

> 1) if an agent is following an object (e.g., cows) to steer it towards a target (e.g., corral), then the state is calculated based on the distance of agent to the object, to the target, and the angle between target, agent and object,
> 2) if an agent is not following an object and just aims to find the target (e.g., goal finding), the state is calculated based on the distance of the agent to the target and the angle between target, agent, and positive side of the x-axis.

The first objective was investigated in the herding problem. For the second objective, our proposed framework is compared with that of [2] in a goal finding environment. A different knowledge fusing mechanism is introduced [2] in which if the distribution of received samples from other teammates are similar, they could be merged. However, in our framework, all the Q-tables related to the same behavior regardless of the similarity of their distributions will be merged by some weights, i.e. number of times pair *(s,a)* was experienced. The presented framework in [2] was evaluated on the Cooperative Goal Searching Problem. In this problem, three cooperative agents are placed in the environment as shown in Fig. 15. The size of the environment is 300 × 300 and it is surrounded by walls and also obstacles. The agents are put in various areas labeled as Area 1, Area 2, and Area 3. Agent 1, Agent 2, and Agent 3 start from their beginning areas R1, R2, and R3 (indicated with the green color in Fig. 15). If an agent finishes an episode, this agent will start randomly from its starting area, which is a 10 × 10 unit, in the following episode. The red area demonstrates the goal. Agent 2 is the nearest to the goal, and it is easier for this agent to reach the goal, while agents 1 and 3 require to explore the environment more extensively to discover a path to avoid the obstacle and reach the desired goal. In other words, in this cooperative task, agent 2 achieves useful knowledge easily, and when agents 1 and 3 enter R2 (Area 2), the knowledge of agent 2 could help the other agents. All three agents have the same action space of {up, down, left, and right}. Throughout the learning process, if the number of steps reached 2000 steps or if the agent reached the goal, an episode gets finished. There were 500 episodes per trial and 40 trials.

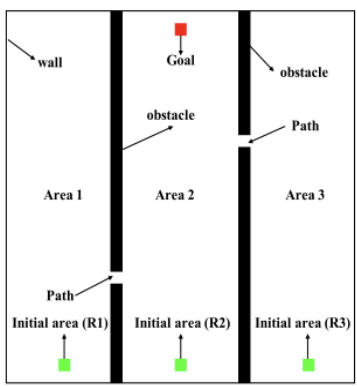

Fig. 15: Cooperative goal search

*We have utilized our proposed framework to solve this problem and compare the results. In our implementation, the state is calculated based on the distance of the agent to the goal and the angle between the goal, agent, and positive side of the x-axis. System roles consist of a Coordinator*

*agent which is an internal agent to the system, as it does not interact with other agents and player agents located in Area1, Area2, and Area3. Three different Q-tables are initialized for each player agent. These Q-tables are filled based on the area of the agent e.g. when agent 1 enters area 3, the Q-table related to area 3 will be filled for this agent. At each time step, all three agents send their Q-table to the coordinator agent. Then, all similar Q-tables( Q-tables related to the same areas) are aggregated by the coordinator agent. The Q-value for the pair s and a in the Q-table is the weighted average of the Q-value in all three agents. Weights are defined as the number of times pairs s and a occur for a specific agent.*

*In Fig. 16-18, a comparison is made between the average number of steps for reaching the goal while using our proposed method and the method introduced in [2]. According to Fig. 16, our proposed method decreased the average steps required for agent 1 to reach the goal down to 130 in the 32$^{th}$ episode. However, this number is 240 for the proposed method in [2] in the same episode. Agent 2 was near the goal; the difference of the average steps was not noticeable as shown in Fig. 17. In Fig. 18, it can be observed that the proposed method decreased the average steps for agent 3 to reach the goal to 300 in the 32$^{th}$ episode. However, this number is 311 for the proposed method in [2] in the same episod*e.Iteration starts from 32, therefore, the difference between numbers on the y-axis can be clearly noticed.

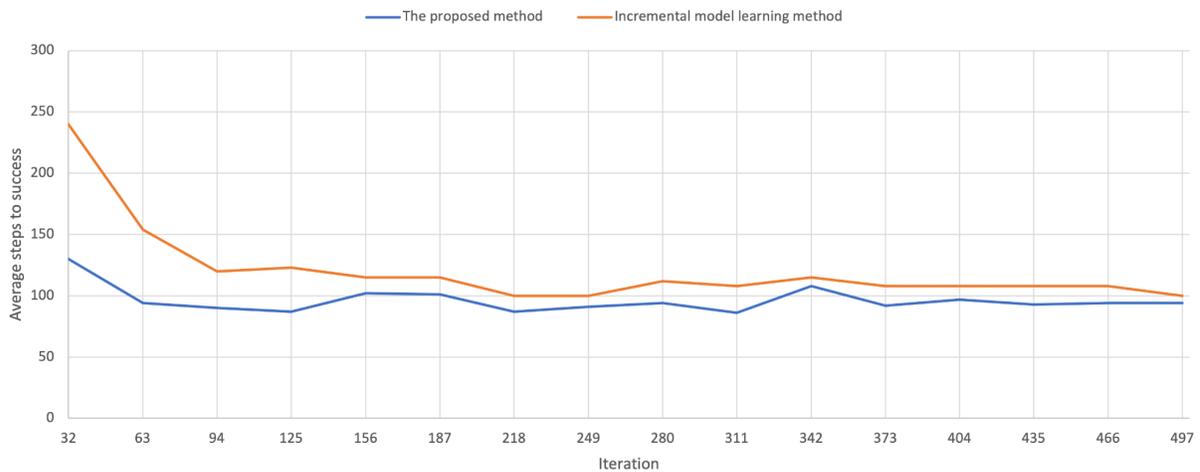

Fig. 16: Average steps to success for Agent 1

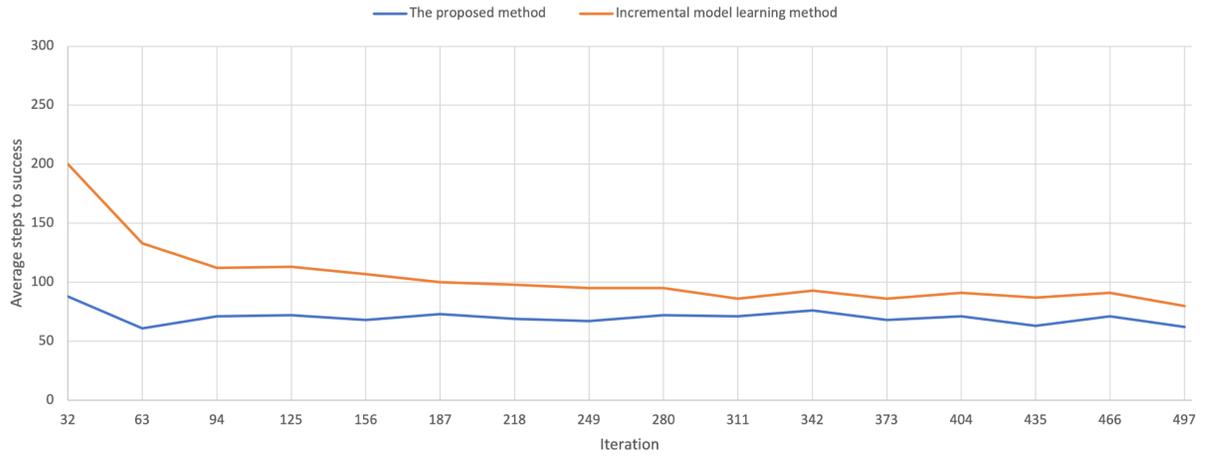

Fig. 17: Average steps to success for Agent 2

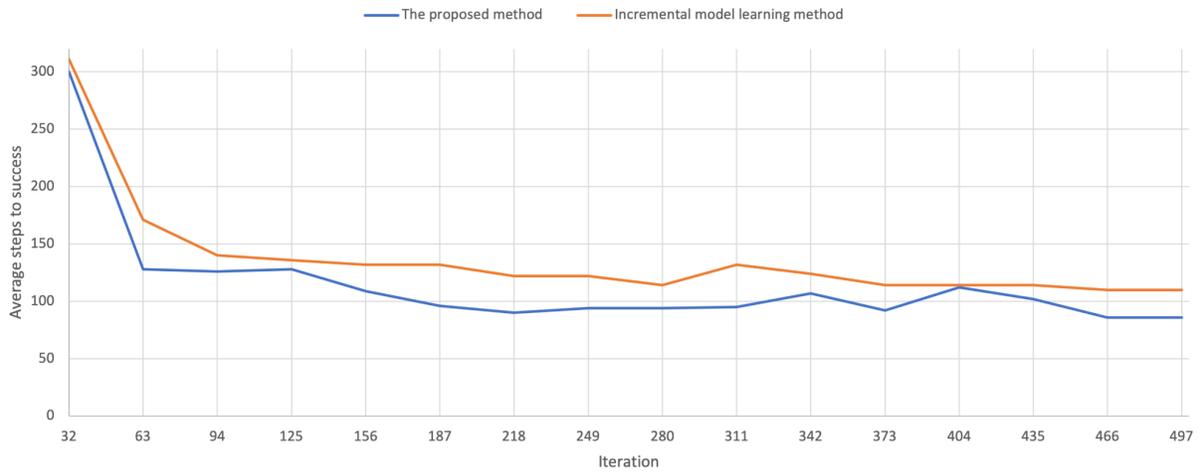

Fig. 18: Average steps to success for Agent 3

A summary of Fig. 16-18 is shown in Tables 4-6. The first column represents the iteration and the second and third columns demonstrate the average steps required for each agent to reach the goal in our proposed method and that of [2] respectively.

Table 4: Average steps required for Agent 1 to reach the goal.

| Iteration | The proposed method | Incremental model learning method |
|---|---|---|
| 32 | 130 | 240 |

| 94 | 90 | 120 |
| --- | --- | --- |
| 218 | 87 | 100 |
| 280 | 94 | 112 |
| 404 | 97 | 108 |
| 466 | 94 | 108 |

*Table 5: Average steps required for Agent 2 to reach the goal.*

| Iteration | The proposed method | Incremental model learning method |
| --- | --- | --- |
| 32 | 88 | 200 |
| 94 | 71 | 112 |
| 218 | 69 | 98 |
| 280 | 72 | 95 |
| 404 | 71 | 91 |
| 466 | 71 | 91 |

*Table 6: Average steps required for Agent 3 to reach the goal.*

| Iteration | The proposed method | Incremental model learning method |
| --- | --- | --- |
| 32 | 300 | 311 |
| 94 | 126 | 140 |
| 218 | 90 | 122 |
| 280 | 94 | 114 |
| 404 | 112 | 114 |

| 466 | 86 | 110 |

## 5. Discussion

Regarding temporal complexity, at the beginning of the implementation without considering the transfer of knowledge between agents, it was assumed that whenever an agent sees its allied agent in its line of sight, it will send a message to start cooperation. In this case, the time complexity is equal to $O(L, M^2)$ in each time step where L is the line of sight and M is the number of agents in a group $1 \ll M \ll N$.

We changed the initial implementation and prevented duplicate collaboration messages by adding a flag. In this case, the temporal complexity is equal to $O(L, M)$ in each time step where $L$ is the line of sight and $M$ is the number of agents in a group $1 \ll M \ll N$.

Then, the coordinator agent is added to the environment as an external agent and the knowledge transfer mechanism is used. At each time step, all agents send their Q-tables to the coordinator agent and the coordinator agent also sends the aggregated Q-table to all agents so the time complexity is equal to $O(L, M+2\times M)$ in each time step where L is the line of sight and $M$ is the number of agents in a group $1 \ll M \ll N$. This complexity could be simplified as $O(N)$ since $M \ll N$.

## 6. Conclusion

In general, solving complex problems has become a major challenge in the field of artificial intelligence. The shepherding problem is one such example due to its large state space as well as a dynamic and uncertain environment. In this paper, we introduce a framework in which if an agent is following an object (e.g., cows) to steer it towards a target (e.g., corral), then the state is calculated based on the distance of agent to the object, to the target, and the angle between target, agent and object. If an agent is not following an object and just aims to find the target (e.g., goal finding), the state is calculated based on the distance of the agent to the target and the angle between target, agent, and positive side of the x-axis. In the proposed framework knowledge is transferred between agents to accelerate the learning process. The new state calculation method declines the size of state space as a state is not calculated based on x-y axis coordinates. This calculation method makes many similar states and also makes the size of Q-tables smaller. The knowledge transferring algorithm decreases the effort required for exploration and declines the learning time in a large environment. Our results demonstrate that the proposed framework significantly reduces the size of the state space by using the concept of state abstraction. The proposed state calculation method increases the number of similar states, thereby increasing the efficiency of the RL algorithm. We also showed that by transferring knowledge between agents, we can reduce the cost of exploration and achieve a higher rate of success. There are several directions for future work. As the number of behaviors increases, it makes sense to train agents to switch between different behaviors automatically. Moreover, implementations could be improved to reduce the messages exchanged between agents.